\documentclass[10pt,letterpaper,twocolumn]{article} 

\usepackage{ol2}

\begin{document}

\twocolumn[

\title{Active control of focal length and beam deflection in a metallic nano-slit array lens with multiple sources}

\author{A. E. \c{C}etin,$^{1}$ K. G\"uven,$^{2,*}$ and \"O. E. M\"ustecapl\i o\~glu$^{2,3}$}

\address{
$^1$School of Electrical and Computer Engineering, Boston University,
Boston, Massachussettes 02215, USA\\
$^2$Department of Physics, Ko\c{c} University, Sar\i yer,
\.Istanbul, 34450, Turkey\\
$^3$Institute of Quantum Electronics, ETH Zurich CH-8093, Switzerland\\
$^*$Corresponding author: kguven@ku.edu.tr}

\begin{abstract}
We propose a surface plasmon-polariton based nano-rod array lens structure that
incorporates two additional lateral input channels, with the ability
to control the focal length and the deflection of the transmitted
beam through the lens actively by the intensity of the channel
sources. We demonstrate by numerical simulations that, applying
the sources with the same intensity can change the focal point and the beam waist,
whereas unequal intensities generate an asymmetric field profile
in the nano-rod array inducing an off-axis beam deflection.
\end{abstract}


 ] 

Metal-dielectric interfaces interacting with light sustain surface plasmon polariton (SPP) excitations that can propagate along the interface. SPPs can provide extraordinarily enhanced light transmission (EOT) through subwavelength apertures in a the metal, beyond the diffraction limited value \cite{cite1,cite2}. Henceforth, there is a growing interest in
developing plasmonic structures for guiding and manipulating the
light propagation at subwavelength scales. Controlling the shape and
direction of the beam emitted through an aperture is one such
feature. Several studies demonstrated the passive control that is
obtained by surrounding the slit with surface corrugations. The
corrugation pattern modifies the SPP dispersion, which, in turn
generates a confined beam in the normal
\cite{cite2,cite3,cite4,cite5}, off-axis \cite{cite6,cite7}, or in
multiple directions \cite{cite8}. The corrugation pattern can also
be modified to modulate the focal length of the beam \cite{cite9, cite10}. Another
design approach employs multiple slits separated by
nanorods without any surface corrugation.  The nanorod profile can
then be utilized for beam shaping \cite{cite11}. Theoretical studies of the surface plasmon-photon interaction on metallic wedge structures provide insight
on the SPP-assisted emission properties\cite{cite12}.

\begin{figure}[t]
\centerline{
\includegraphics[width=8.3cm]{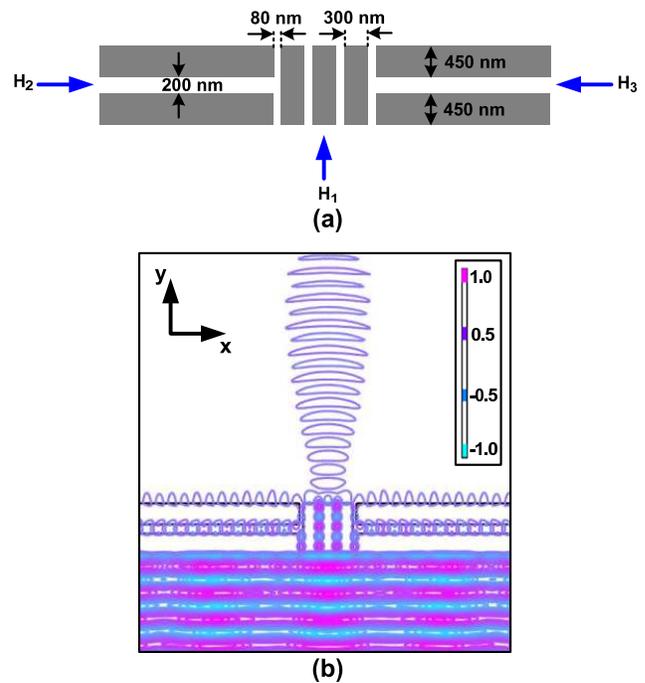}}
\caption{(a) Schematic of the nano-rod lens system with the main input field $H_{1}$ and two control sources,
$H_{2}$, $H_{3}$. (b) Simulation of the
system at $H_{1}=1$ [V/m] and $H_{2}=H_{3}=0.2$ [V/m].}
\label{figure1}
\end{figure}

Evidently, active control of the beam modulation is highly
desirable. A recent study proposes embedding
Kerr nonlinear medium in the slit-array, where the nonlinearity
driven by the intensity of the incident beam induces beam deflection
and focusing upon inducing a specific phase retardation at each slit
\cite{cite13}. This elegant approach presents promising results, but
filling the nanoslits with a nonlinear medium is challenging.

In this paper we propose a new two-dimensional nano-optic lens structure
consisting of multiple metallic nano-rods that are fed through a main (input)channel and two lateral (control) channels. By harnessing the SPPs,
both the focal length and the deflection of the transmitted beam through the slit-array
can be controlled actively by the intensity of the channel sources
without requiring any nonlinear response.

The lens structure is depicted in Fig. \ref{figure1}(a). Transverse-magnetic (TM) polarized monochormatic waves are propagating through the main input ($H_{1}$) and the lateral control channels ($H_{2}, H_{3}$). Field amplitudes are normalized with respect to that of the main input. $H_{2}, H_{3}$ are incident to the outermost nano-slits of the lens. The field propagations are simulated by the Comsol Multiphysics software which employs the finite element method along with time-domain calculation of the fields. The metal is modeled by the dielectric function of silver at $\lambda=561$ nm $\varepsilon=-11.66+i0.3771$. Unless noted otherwise, the figures depict the magnetic field component in contour plots. Figure \ref{figure1} (b) shows a snapshot of the field when all the sources are active.

\begin{figure}[htpt]
\centerline{
\includegraphics[width=8.3cm]{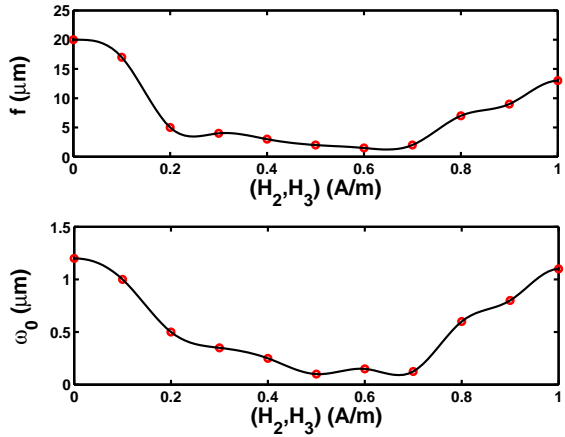}}
\caption{(Color online) Focus length, $f$ and beam waist $\omega_{0}$ of the
transmitted beam as a function of the identical amplitude of the control
sources, $H_{2}$ and $H_{3}$.}
\label{figure2}
\end{figure}

We begin by describing the control of focal length and beam waist via fields applied with equal
amplitudes in the lateral channels. Figure \ref{figure2}, shows that the emitted beam is collimated as the amplitudes of $H_{2}$ and $H_{3}$ are increased up to 0.5 [V/m]. The sensitivity of control is fine when $H_{2}$,$H_{3}$ are much less than $H_{1}$, and becomes coarse between 0.3 - 0.5 [V/m]. When $H_{2}$,$H_{3}$ start to become comparable to $H_{1}$, defocusing occurs.

The behavior can be described qualitatively by the effect of SPPs generated at the outermost nanorod surfaces. In the regime $H_{2}$, $H_{3}$ << $H_{1}$, the control fields have negligible direct contribution to the transmitted field, but the SPPs induced by them interfere with that of the main input field, reducing the transmission through the outermost slits. When the outermost slits become dimmer with increasing control field amplitude, the effective aperture of the lens is reduced to the innermost slits. Consequently, both the focal length and the beam waist decreases. When $H_{2}$,$H_{3}$ become comparable to $H_{1}$, the outermost slits become bright in transmission and the transmitted beam appears diffracted. In loose terms, the lateral channels can change selectively the number and location of "on" slits in the lens. The resulting behavior is consistent with that in Ref. \cite{cite11} where the number of slits were changed structurally, and the increase of directivity and focal length was observed with increasing slit number. The role of SPPs were also demonstrated in the context of interference effects in double and multiple metallic slit geometries. \cite{cite14,cite15,cite16} It is shown that the SPPs excited at the slit surfaces propagate phase information between the slits and affect the transmission. Ref. \cite{cite16} reports that the mutual interference of SPPs takes precedence in determining the transmission properties in the case of a multislit structure. In accordance with these results, we demonstrate that modifying the SPPs in the multislit lens structure can be utilized as a beam control mechanism.

\begin{figure}[htpt]
\centerline{
\includegraphics[width=8.3cm]{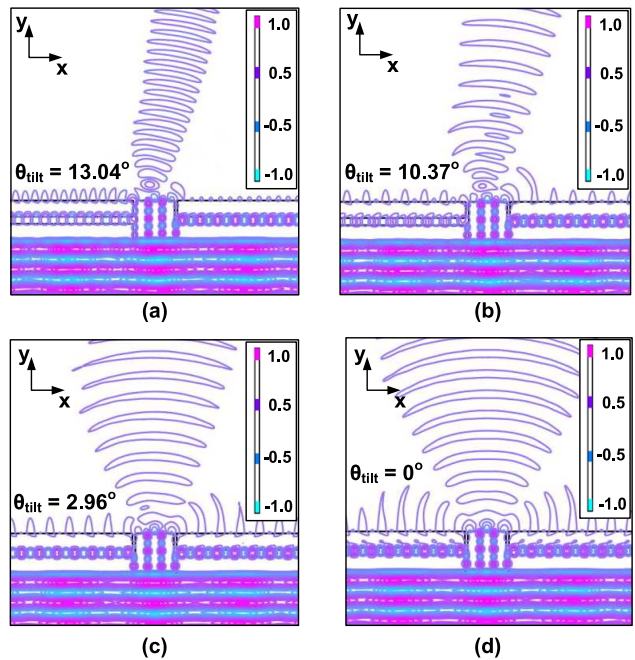}}
\caption{(Color online) Beam deflection by the nano-rod array lens at fixed $H_{1} = 1$ [V/m] and $H_{3} = 0.5$ [V/m] for the value of $H_{2}$ = (a) 0 [V/m] (b) 0.2 [V/m] (c) 0.4 [V/m] and (d) 0.5 [V/m]}
\label{figure3}
\end{figure}

\begin{figure}[htpt]
\centerline{
\includegraphics[width=8.3cm]{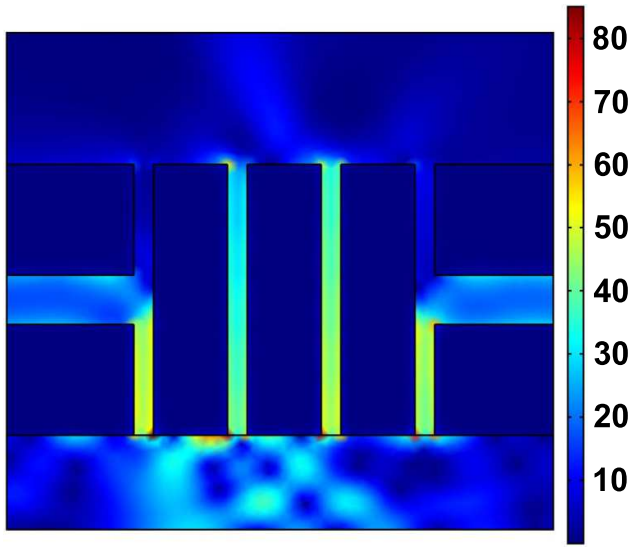}}
\caption{(Color online) The field intensity in the nano-rod array
for $H_{1}=1.0$, $H_{2}=0.2$ and $H_{3}=0.5$ [V/m] corresponding to Fig. \ref{figure3}(b)}
\label{figure4}
\end{figure}

With the ability to control the SPPs on the nano-rods of the lens, one can apply unequal field amplitudes to generate an asymmetric SPP profile. In Figure \ref{figure3}, we demonstrate that this induces a tilting of the
transmitted beam. In this simulation, the amplitude of $H_{2}$ is varied between 0.0-0.5 [V/m] while $H_{1} = 1$[V/m] and $H_{3} = 0.5$ [V/m] are kept fixed. The tilting is evident in Fig.s \ref{figure3} (a) and (b) corresponding to $H_{2} << H_{3}$. The asymmetric field intensity distribution within the nano-rod lens for the source settings of Fig. \ref{figure3} (b) is plotted in Fig. \ref{figure4}. When $H_{2}$ is increased, the tilt angle decreases with the reduced asymmetry of the field profile and vanishes completely for the symmetric field distribution (Fig.s \ref{figure3} (c), (d)). But this is accompanied by the defocusing effect, since $H_{2}$ and $H_{3}$ are now comparable to $H_{1}$.

\begin{figure}[htpt]
\centerline{
\includegraphics[width=8.3cm]{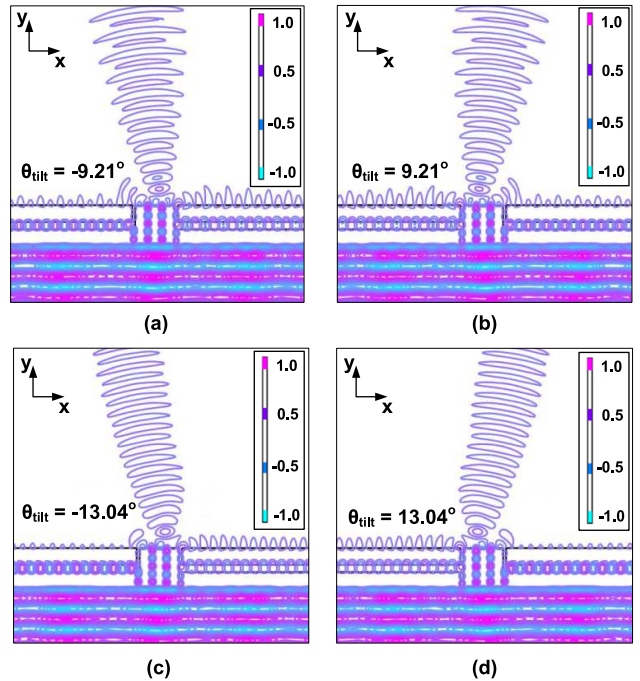}}
\caption{(Color online) Discrete beam scan operation by switching the control sources
alternatingly.(in [V/m] units)
(a) $H_{2}=0.3$, $H_{3}=0$ (b) $H_{2}=0$, $H_{3}=0.3$ (c) $H_{2}=0.5$, $H_{3}=0$ (d) $H_{2}=0$,
$H_{3}=0.5$.}
\label{figure5}
\end{figure}

By adjusting the amplitude of the control sources properly, it may be possible to scan the beam while preserving its shape. Figure \ref{figure5}, demonstrates a discrete beam scan operation with the control sources are being switched alternatingly. A continuous beam scan ability is limited, since the focal length and beam waist depend on the intensities of the control sources.

In conclusion, we propose a SPP based nano-rod array lens system with dual control sources to manipulate the optical beam emitted through a subwavelength aperture actively. The beam profile depends on the relative intensity and phase of the field in each slit which, in turn, depends on the spp excitation in the slit. This is roughly analogous to a phased array antenna system. We define two control modes: (I)Symmetric intensity change, that turns the outermost slits "on" and "off" in relative intensity with respect to the innermost slits. This modifies the focal length and directivity since the number and position of "on" slits is changing. (II) Asymmetric intensity change in the lateral control channels tilts the beam, as now both the number and horizontal location of the slit group are changing.

\section*{Acknowledgments}This work is supported by Technological Research Council of Turkey under research grants and project no. 106E198, and Turkish Academy of Sciences GEBIP.

\end{document}